\documentclass[showpacs,preprintnumbers,amsmath,amssymb,prl,twocolumn]{revtex4}

\usepackage{amsmath}
\usepackage{amssymb}
\usepackage[dvips]{graphicx}
\usepackage[dvips]{color}
\usepackage[T2A]{fontenc}
\usepackage[cp1251]{inputenc}
\usepackage[english]{babel}

\newcommand{\bra}[1]{\left\langle#1\right\vert}
\newcommand{\ket}[1]{\left\vert#1\right\rangle}

\begin{document}

\title{Angular Schmidt Modes in Spontaneous Parametric Down-Conversion}
\author{S.S.Straupe, D.P.Ivanov, A.A.Kalinkin, I.B.Bobrov, S.P.Kulik}
\affiliation{Faculty of Physics, M.V.Lomonosov Moscow State University, 199001, Moscow, Russia}

\date{\today}
\begin{abstract}

We report a proof-of-principle experiment demonstrating that appropriately chosen set of Hermite-Gaussian modes constitutes a Schmidt decomposition for transverse momentum states of biphotons generated in the process of spontaneous parametric down conversion. We experimentally realize projective measurements in Schmidt basis and observe correlations between appropriate pairs of modes. We perform tomographical state reconstruction in the Schmidt basis, by direct measurement of single-photon density matrix eigenvalues.

\end{abstract}
\maketitle


\noindent\textbf{Introduction}.
Characterizing entanglement in high-dimensional quantum systems is a hot topic in quantum information science. One of the less studied subjects is entanglement in the extreme case of infinite-dimensional by-partite states. It is, nevertheless, of much interest, since such states are usual for two-particle scattering processes, such as spontaneous parametric down conversion (SPDC), being one of the most experimentally attractive source of entangled photons. An approach to quantifying entanglement in this case was proposed by Law and Eberly \cite{EberlyPRL04}. It is based on \emph{Schmidt decomposition} - a representation of a bipartite state vector as a sum of factorized terms: $\ket{\Psi_{12}}=\sum_k{\sqrt{\lambda_k}\ket{u_k}\ket{v_k}}$, with $\ket{u_k},\ket{v_k}$ being eigenvectors of reduced single-particle density matrices (so called Schmidt modes) and $\lambda_k$ - corresponding eigenvalues. This decomposition has several remarkable features:
\begin{itemize}
\item{\noindent Schmidt modes, being pure single-particle states, form a complete and orthogonal basis;}
\item{\noindent decomposition has a discrete single-sum form, i.e. every Schmidt mode is correlated with exactly one counterpart;}
\item{\noindent number of significant eigenvalues, defined as $K=1/\sum_k{\lambda_k^2}$}, quantifies entanglement present in the system.
\end{itemize}
In this Letter we experimentally address the physical properties corresponding to the listed features of Schmidt decomposition. Specifically, we choose spatial degrees of freedom of biphoton field as an experimentally convenient example of an infinite-dimensional system.

Angular spectrum of biphotons is well recognized as an attractive object to study multidimensional entanglement and its properties were experimentally investigated in numerous works during the last decade. These were focusing on two main alternatives: EPR-like correlations in transverse momentum \cite{HowellPRL04, HowellPRL, PaduaPRL, Fedorov} and entanglement in orbital angular momentum \cite{ZeilingerNature01, PadgettPRL02, MonkenPRA04, VaziriPRL02, VaziriPRL03, TornerPRL, LangfordPRL04, WoerdmanPRL05, PadgettPRL10, vanExterPRA07, vanExterPRL10, BoydPRL10}. Both approaches may be addressed on the same grounds: transverse entanglement manifests itself in non-classical correlations between coherent spatial modes - plain waves in the first case, and arbitrarily chosen Laguerre-Gaussian modes in the second one. None of these choices is perfect in the following sense: the correlations between modes are non-ideal, meaning that a single mode is correlated to multiple counterparts. For EPR case the number of correlated modes is limited from below by finite angular divergence of the pump, making the biphoton state always less entangled then an ideal EPR-pair \cite{HowellPRL04, Fedorov}. In the case of OAM-entanglement, LG modes with different radial indices are in general correlated \cite{LG-schmidt}. It is exactly the Schmidt decomposition providing a natural set of modes to study transverse entanglement in SPDC, meaning that, apart from fundamental interest, it is of interest for high dimensional quantum state engineering.

\noindent\textbf{Angular Schmidt Modes}.
Biphoton state generated in type-I SPDC process has the following form \cite{Klyshko}: $\ket{\Psi}=\ket{vac}+\int{d\vec{k_1}d\vec{k_2}\Psi(\vec{k_1},\vec{k_2})\ket{1}_{k_1}\ket{1}_{k_2}}$, with $\vec{k_{1,2}}$ being the wavevectors of scattered photons. In the thick-crystal approximation and neglecting the walk-off effect caused by birefringence of the nonlinear crystal, the biphoton amplitude $\Psi(\vec{k_1},\vec{k_2})$ is described by the following expression \cite{BurlakovPRA97, MonkenPRA98, MonkenPRA03}:
\begin{equation}\label{Monken}
    \Psi(\vec{k_1},\vec{k_2})\propto\mathcal{E}_p(\vec{k_{1\perp}}+\vec{k_{2\perp}})\mathrm{sinc}\left[\frac{L(\vec{k_{1\perp}}-\vec{k_{2\perp}})^2}{4k_p}\right],
\end{equation}
where $\mathcal{E}_p(\vec{k})$ stands for angular spectrum of the pump, $L$ is length of the crystal, $k_p$ - wavevector of the pump, and subscript $\perp$ denotes transverse vector component. We may rewrite the amplitude in a form of Schmidt decomposition:
\begin{equation}\label{Schmidt-decomposition}
    \Psi(\vec{k_{1 \perp}},\vec{k_{2 \perp}}) = \sum_{i=0}^{\infty} {\sqrt{\lambda_{i}}\psi_i(\vec{k_{1\perp}})\psi_i(\vec{k_{2\perp}})}.
\end{equation}

Unfortunately, there is no analytical expression for Schmidt modes known for an exact wavefunction (\ref{Monken}). We will assume the pump to be gaussian and approximate the biphoton amplitude by a double-gaussian form as proposed in \cite{WalmsleyQIC03, FedorovJPB09}:
\begin{equation}\label{double-gaussian}
    \begin{split}
    &\Psi(\vec{k_1},\vec{k_2})\propto \\ &\exp\left(-\frac{(\vec{k_{1\perp}}+\vec{k_{2\perp}})^2}{2a^2}\right)\exp\left(-\frac{(\vec{k_{1\perp}}-\vec{k_{2\perp}})^2}{2b^2}\right),
    \end{split}
\end{equation}
which allows one to find Schmidt decomposition in an analytical form \cite{EberlyPRL04}. The wavefunction (\ref{double-gaussian}) depends on two experimentally adjustable parameters: pump divergence $a$ and phase-matching angular width $b$, and the corresponding Schmidt modes may be shown to be Hermite-Gaussian or Laguerre-Gaussian modes with appropriately chosen widths. Namely, in cartesian coordinates $\{k_{1(2)x},k_{1(2)y}\}$ the decomposition reads:
\begin{equation}\label{HG-Schmidt-2D}
    \begin{split}
    &\Psi(\vec{k_1},\vec{k_2})= \\ &\sum_{mn}{\sqrt{\lambda_n\lambda_m}\psi_n(k_{1x})\psi_m(k_{1y}) \times \psi_n(k_{2x})\psi_m(k_{2y})},
     \end{split}
\end{equation}
with $\psi_n(k_{1x,2x})=\left(\frac{2}{ab}\right)^{1/4} \phi_n \left(\sqrt{\frac{2}{ab}} k_{1x,2x}\right)$ and $\phi_n(x)= (2^n n!\sqrt{\pi})^{-1/2}e^{-x^2/2}H_n(x)$ being Hermite-Gaussian functions. Eigenvalues decrease exponentially with growing $n$:
\begin{equation}\label{HG-eigenvalues}
    \lambda_n = 4ab\frac{(a-b)^{2n}}{(a+b)^{2(n+1)}}.
\end{equation}
We performed the numerical calculation of eigenvalues and eigenmodes for exact wavefunction (\ref{Monken}) for the experimental values of pump and phase-matching bandwidths. The results were indeed close to those predicted by double-gaussian model, validating its applicability in our experimental conditions.

Since expression (\ref{HG-Schmidt-2D}) is a product of two decompositions depending on $k_{1,2x}$ and $k_{1,2y}$, the Schmidt number, quantifying entanglement in the system, has the form $K=K_x\times K_y$, with:
\begin{equation}\label{Schmidt-number}
    K_x = K_y = \left(\frac{1}{\sum_n^\infty{\lambda_n^2}}\right) = \frac{a^2+b^2}{2ab}.
\end{equation}

\noindent\textbf{Experimental realization}.
Our main goal was to realize projective measurements in Schmidt basis experimentally. The simplest case is a zeroth-order $HG_{00}$ mode, which can be filtered by appropriate coupling to a fundamental gaussian mode of the single-mode fiber. So a single-mode fiber followed by a photon-counting detector would realize a projector on the $HG_{00}$ mode. For higher order modes one has to use phase holograms, transforming the gaussian beam into $HG_{nm}$ mode. This transformation may be realized with high efficiency using phase only holograms \cite{WoerdmanOptComm94, PadgettOptCom96, WegenerOptQuantEl92}. When the appropriate hologram is chosen, the corresponding $HG$ mode is transformed into a Gaussian one, whereas the others transform to orthogonal ones, so only that mode is transmitted through the fiber, realizing a desired projection. For Hermite-Gaussian modes the holograms look essentially like step-wise phase masks introducing a phase shift equal to $\mathrm{Arg}(H_n(x)H_m(y))$ into the beam.

\begin{figure}[!h]
\centering\includegraphics[width=0.5\textwidth]{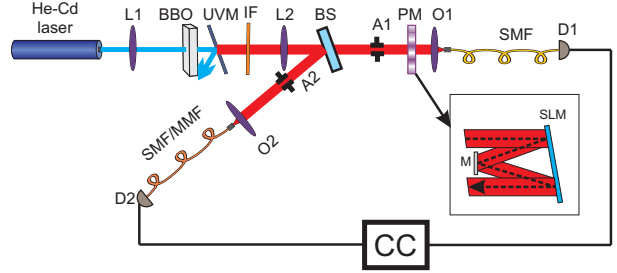}
\caption{{\protect{Experimental setup realizing projective measurements in Schmidt basis (see text for details). Phase mask PM is shown as transparent for simplicity, actual realization uses double reflection from SLM as shown on the inset.}}} \label{setup}
\end{figure}

Schematic of experimental setup is presented in Fig.\ref{setup}. We used a 325 nm CW He-Cd laser to pump a 2mm thick BBO crystal cut for collinear frequency degenerate Type-I phase-matching. The phase-matching bandwidth was estimated to be $b = 20$ mrad. Laser beam was focused into the crystal by a 150 mm quartz lens L1, leading to divergence $a = 5.8$ mrad. The particular values of parameters were chosen to reduce the Schmidt number to a reasonably low value of $K\sim10$, so that only several modes of lower order were significant. Ultraviolet mirror UVM and wide interference filter IF with 40 nm bandwidth were used to cut off the pump and additional luminescence. The crystal was put in the focal plane of a 140 mm lens L2 to collimate the SPDC radiation into a $\sim3$ mm wide beam which, after being split on a 50/50 non-polarizing beam-splitter BS, was coupled with 8X microscope objectives O1,2 to single-mode fibers SMF placed in objectives' focal planes. The optical setup was aligned to match the fundamental mode of the fiber to a gaussian beam corresponding to zeroth-order Schmidt mode for SPDC radiation. The fibers were connected to single-photon counting modules (Prekin-Elmer), followed by a coincidence circuit.

We used a reflective Liquid-Crystals-on-Silicon phase-only SLM placed in the transmitted arm of the BS to display the holograms. The device used has 1280x768 square pixels of $10 \mu m$ size. To check the quality of transformation between HG modes, we used a 650 nm single-mode fiber coupled diode laser to produce a gaussian beam, which was expanded by a 20X microscope objective to match the SPDC mode size. That was a coherent source similar to $HG_{00}$ Schmidt mode. The holograms were adjusted to minimize the counts rate for a transformed beam, ensuring its orthogonality to the fundamental gaussian mode. We obtained minima with visibility greater than 97\% for modes with $n,m\leq4$. We further checked shapes of the modes obtained by scanning the detecting SMF tip in the focal plane of the coupling objective. The obtained dependencies of counting rate on fiber tip position $R(x)$ showed behavior exactly similar to expected convolution $R(x) \propto \int_{-\infty}^{\infty} H_{nm}(\tilde{x}/a)\exp\left({-\frac{\tilde{x}^2}{a^2}}\right)\exp\left(-\frac{(x-\tilde{x})^2}{a^2}\right)$.

\begin{figure}[!h]
\centering\includegraphics[width=0.5\textwidth]{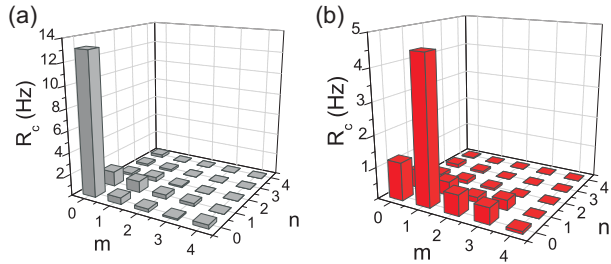}
\caption{{\protect{Coincidence counting rates for $HG_{mn}$ modes selected in the transmitted arm of the setup, while $HG_{00}$ mode (a) and $HG_{10}$ (b) modes are selected in the reflected arm.}}} \label{visibility}
\end{figure}

Similar behavior  observed for SPDC radiation when $HG_{00}$ mode was selected in the reflected arm, while one of the orthogonal $HG_{nm}$ modes was selected in the transmitted arm. Coincidence counts rate $R_c$ for holograms with different $n$ and $m$ in the transmitted arm are shown in Fig.{\ref{visibility}}(a). Only $n=m=0$ case displays high counting rate, while all others are suppressed with visibility over 90\%, in good agreement with what is expected for Schmidt decomposition. When some higher order mode is selected in the reflected arm also, coincidences are expected to be registered only when its counterpart is selected in the transmitted arm, which is confirmed by experimental results presented in Fig.\ref{visibility}(b). We observe the somewhat lower visibility in this case, which may be explained by low quality of phase masks used in the reflected channel, since phase-step plates made of glass were used instead of the SLM in that arm of the setup. We believe the anomaly higher counting rates for $HG_{01}$ and $HG_{11}$ modes to have their origin in the fact that alignment in the vertical plane was much more coarse than in the horizontal one, due to technical reasons. These results illustrate the single-sum property of Schmidt decomposition, i.e. demonstrate the pairwise correlations being characteristic for Schmidt modes. 

\begin{figure}[!h]
\centering\includegraphics[width=0.5\textwidth]{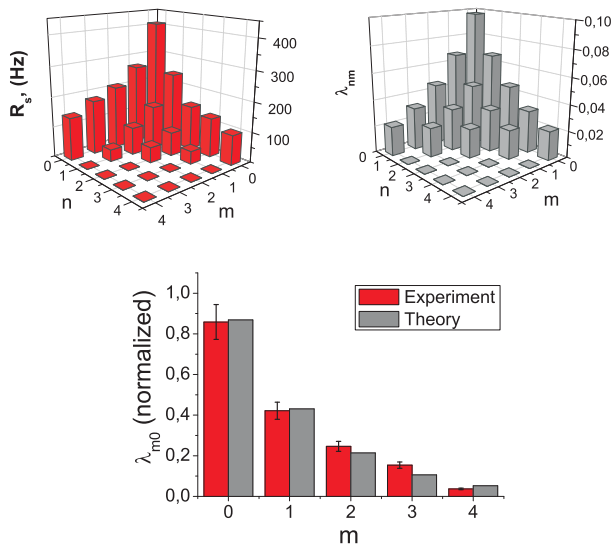}
\caption{{\protect{(a) Single counts rates for $HG_{mn}$ modes selected in the transmitted arm of the setup. Not normalized, only background noise was subtracted. (b) Single-photon density matrix eigenvalues $\lambda_{mn}$ in HG-modes basis, which are the weights in the Schmidt decomposition. (c) One-dimensional projection on the plane $n=0$. Red bars are normalized experimental results, black bars - results of numerical calculation with exact wavefunction (\ref{Monken}), blue curve - eigenvalues for HG-modes predicted by double-gaussian model.}}} \label{eigenvalues}
\end{figure}

When $HG_{nm}$ mode and $HG_{00}$ modes were selected in the transmitted and the reflected arms, respectively, and the fiber tip was scanned, the coincidence rates behave like an expected convolution of corresponding modes. Moreover, absolutely similar behavior was observed, irrespectively of whether the fiber in the transmitted or in the reflected arm was scanned. This behavior is a direct experimental evidence of the fact, that projecting one of the photons into a Schmidt mode leads to a pure (spatially coherent) state for the second one, which is another feature specific for Schmidt decomposition. At the same time a very different dependencies were observed for single counts of the detector in the transmitted arm - smooth wide curves with no local minima, as we would indeed expect for spatially multi-mode radiation. The value of single counts rate for central positions of both fiber tips is directly related to single-photon density matrix eigenvalues in Schmidt basis: $R_s \sim \bra{\psi_{mn}}\rho^{(s)}\ket{\psi_{mn}} \sim \lambda_{mn},$ providing a direct way to perform state tomography. Results for this kind of measurements are presented in Fig.\ref{eigenvalues}(a). These should be compared with eigenvalues calculated according to the analytical expression (\ref{HG-eigenvalues}) illustrated in Fig.\ref{eigenvalues}(b). To make the comparison quantitative, one may use Uhlmann's fidelity
\begin{equation}\label{fidelity}
    F= \mathrm{Tr}\sqrt{\sqrt{\rho}\rho^{(exp)}\sqrt{\rho}} = \sum_{m,n}{\sqrt{\lambda_{mn}\lambda_{mn}^{(exp)}}},
\end{equation}
where $\lambda_{mn}^{(exp)}$ are experimental estimates for eigenvalues obtained by appropriate normalization of single counts rates. We achieved a value of $F=(92\pm3)\%$, showing a good agreement with double-gaussian model.

Experimental measurement of eigenvalues in Schmidt decomposition provides a direct way of quantifying entanglement in considered system. We have achieved the following experimental estimates for Schmidt numbers: $K_x = 3.1\pm0.9$ and $K_y=2.7\pm0.5$ in agreement with theoretical prediction of $K_{x,y}=2.97$ obtained from (\ref{Schmidt-number}).

\begin{figure}[!h]
\centering\includegraphics[width=0.5\textwidth]{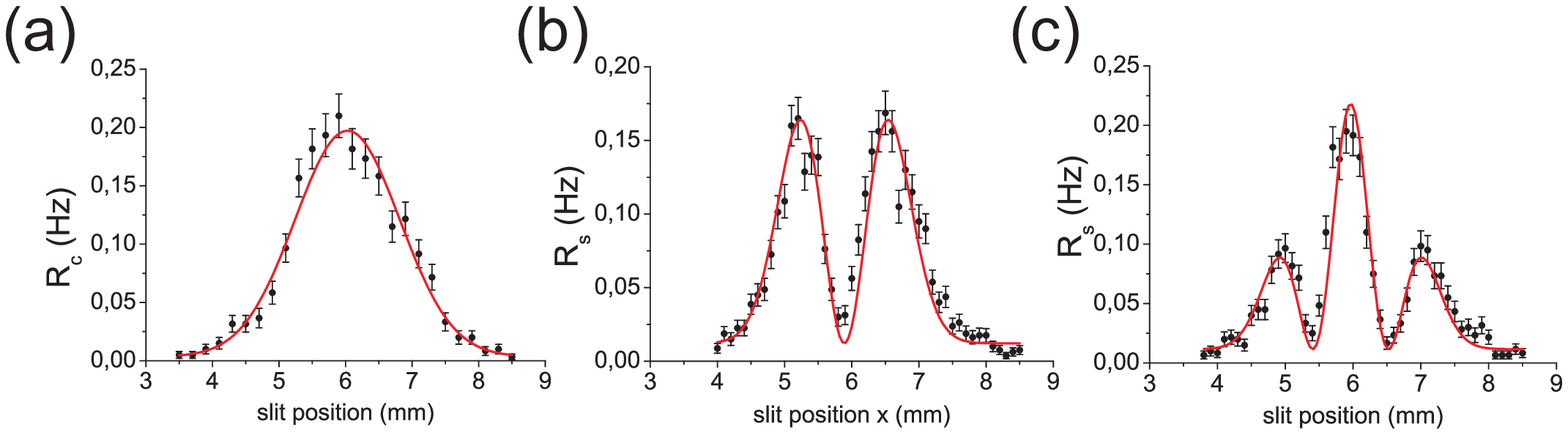}
\caption{{\protect{Ghost images of the first three Schmidt modes: $HG_{00}$ (a), $HG_{10}$ (b) and $HG_{20}$ (c), correspondingly. Solid red curves are Hermite-Gaussian fits.}}} \label{ghost}
\end{figure}

The most direct measurement of Schmidt modes shape may be performed with ghost-interference techniques \cite{ShihPRL95}. For this purpose we substituted the single-mode fiber in the reflected channel with a multi-mode one with approximately ten times larger core diameter. Multi-mode fiber, supporting many spatial modes, served as a "bucket" detector, collecting the entire angular spectrum of SPDC radiation. A 200 $\mu m$ slit was scanned in front of the focusing objective in the same arm to obtain an image in the coincidence counts. In this case, we can directly resolve the angular shape of the desired mode with corresponding hologram inserted in the transmitted arm. Fig.\ref{ghost} shows the obtained "ghost" interference patterns. They are well approximated by Hermite-Gaussian functions, as expected.

\noindent\textbf{Discussion}.
We have reported here the first, up to our knowledge, experimental attempt to address the physical properties of Schmidt decomposition for an infinite-dimensional system. We have applied the classical technique of spatial modes transformation to study transverse entanglement in SPDC in Schmidt basis. Although a significant amount of work was devoted to orbital angular momentum entanglement of Lagguerre-Gaussian modes, the question of conditions under which they constitute a Schmidt basis is much less studied. Results presented here demonstrate, that another basis, namely that of Hermite-Gaussian modes, may be equally attractive for applications, although it is surprisingly much less used in spatial entanglement experiments. Cartesian coordinates on the plane of transverse momentum components of photons, corresponding to HG-modes, may be especially convenient in the case of high transverse entanglement, when the biphoton amplitude is anisotropic, having different shapes in direction of optical axis of the crystal and in orthogonal one \cite{Fedorov}.

Presented results clearly show, that (at least for mildly focused pump) a set of Hermite-Gaussian modes has all the properties of a Schmidt decomposition. Namely, perfect one-to-one correlations between modes, spatial coherence and expected behavior of eigenvalues.

Besides studying spatial entanglement in SPDC itself, Schmidt decomposition provides a natural way to get pure spatial states of heralded single photons, which is of great importance for quantum information tasks. It is thus an important step to demonstrate the realization of projective measurements in Schmidt basis. An interesting theme to address is the possibility to prepare an initially factorized spatial state of biphoton pairs by manipulating the angular spectrum of the pump. Development of Schmidt mode filtering techniques in frequency domain would also be an interesting challenge. We believe that proof-of-principle results of this work to be another step on the way to mastering high dimensional quantum state engineering with spatial states of photons.

We are grateful to M.V. Fedorov for stimulating discussions.

This work has been funded by Ministry of education and science of the Russian Federation (Minobrnauka), the state contract 02.740.11.0223 and by the RFBR grants 10-02-00204 and 10-02-90036\_Bel. S.S.Straupe is grateful to the Dynasty Foundation for financial support.


\begin{thebibliography}{10}
\bibitem{LG-schmidt}
LG-modes may constitute a Schmidt decomposition if the gaussian width is appropriately chosen, according to the relative width of pump angular spectrum and phasematching bandwidth, as described later in this manuscript for HG-modes.
\end{thebibliography}

\end{document}